\documentclass[11pt,letterpaper]{article}
\usepackage[utf8]{inputenc}
\usepackage[english]{babel}
\usepackage{amsmath}
\usepackage{amsfonts}
\usepackage{amssymb}
\usepackage{graphicx}
\usepackage[left=3cm,right=3cm,top=2cm,bottom=3cm]{geometry}
\author{Jorge Pinochet}
\title{\textbf{“Black holes ain't so black”: An introduction to the great discoveries of Stephen Hawking}}
\begin{document}

\author{Jorge Pinochet\\ \\
 \small{\textit{Facultad de Educación}}\\
 \small{\textit{Universidad Alberto Hurtado, Erasmo Escala 1825, Santiago, Chile.} japinochet@gmail.com}\\ \\}

\date{\small \today}
\maketitle

\begin{abstract}
\noindent Between 1974 and 1975, Stephen Hawking revolutionized the world of physics by proposing that black holes have temperature, entropy, and evaporate gradually. The objective of this article is to offer a brief and updated introduction to these three remarkable results, employing only high school algebra and elementary physics. This article may be useful as pedagogical material in an introductory undergraduate physics course.\\ \\

\noindent \textbf{Keywords}: Black holes, Hawking discoveries, introductory undergraduate physics course. 

\end{abstract}

\maketitle

\section{Introduction}

The recently deceased Stephen Hawking was one of the greatest scientists of the second half of the 20th century. According to specialists, his most important contribution to physics was his theoretical demonstration that "black holes ain't so black" [1]; in other words, Hawking discovered that black holes emit radiation as if they were hot bodies. Three direct consequences of this discovery are that black holes have a non-zero temperature, possess entropy, and evaporate gradually. The objective of this article is to offer a brief and updated introduction to these three remarkable results, employing only high school algebra and elementary physics.\\

The article is organized as follows. In Sections 2, 3 and 4, I separately analyze Hawking's findings. Section 2 also serves as a brief introduction to the concept of black hole. Finally, in section 5 I discuss the possibility of empirically corroborating Hawking's findings.

\section{Hawking radiation and Hawking temperature}
The classic black hole paradigm, accepted almost without question until the beginning of the 1970s, was based only on general relativity, which was the theory of gravity proposed by Einstein in 1916 to extend and perfect the law of Newtonian gravity [2]. According to this paradigm, a black hole is a region of space-time that is limited by a closed surface called the \textit{horizon}, within which there is such a high concentration of mass-energy that nothing can escape its gravity, not even light. In the simplest case, corresponding to a spherically symmetric black hole without rotation and electrically neutral, called \textit{static black hole}, the horizon can be intuitively visualised as a spherical surface whose radius only depends on the mass:

\begin{equation} 
 R_{S} = \dfrac{2GM_{BH}}{c^{2}} = 1.48\times 10^{-27} m \left( \dfrac{M_{BH}}{kg} \right). 
 \end{equation} 

This is the \textit{gravitational radius}\footnote{In the intuitive framework of Newtonian gravitation, $R_{S}$ can be imagined as the distance separating the center of the black hole (where the so-called singularity is) from the horizon; nevertheless, in general relativity $R_{S}$ is not a physical distance but a coordinate.}, where $G= 6.67 \times 10^{-11} N\cdot m^{2}\cdot kg^{-2}$ is the universal gravitation constant, $c= 3\times10^{8} m\cdot s^{-1}$  is the speed of light in vacuum, and $M_{BH}$ is the mass of the black hole. According to the classical paradigm, in order to pass the horizon from the interior to the exterior, a speed is required that is greater than $c$, which is forbidden by relativistic physics. Hence, no type of radiation can be emitted from the horizon, which implies that its temperature must be strictly zero. However, this paradigm changed dramatically in 1974, when Hawking combined general relativity with quantum mechanics to study black holes. As a result, Hawking made the theoretical discovery that an isolated static hole emits thermal radiation in all directions, with a black body spectrum with an absolute temperature given by\footnote{The fact that these three results are applied to isolated black holes is an important aspect of Hawking's finding, since it means that the emission of radiation does not depend on mechanisms related to the presence of material on the outskirts of the horizon, as happens with the accretion, which is a physical process that generates large radiation emissions in black holes that are part, for example, of binary systems.} [3]:

\begin{equation}	
T_{H}= \dfrac{\hbar c^{3}}{8\pi kGM_{BH}} = 1.23\times 10^{23} K \left( \dfrac{kg}{M_{BH}}\right), 
\end{equation}

where $\hbar = h/2\pi = 1.05\times 10^{-34} J\cdot s$ is the reduced Planck constant, and $k= 1.38\times 10^{-23} J\cdot K^{-1}$ is the Boltzmann constant. $T_{H}$ is known as the \textit{Hawking temperature} \footnote{An important but subtle point about this result is that $T_{H}$ is not the temperature near the horizon, but rather it is the temperature measured at a great distance (ideally infinite).}, and the associated thermal emission is called \textit{Hawking radiation}. Eq. (2) was introduced by Hawking in two articles that have become classics of theoretical physics [4, 5] (this equation also appears in Hawking's popular science book, \textit{The Universe in a Nutshell} [6]). In these articles, Hawking also introduces the concepts of evaporation and black hole entropy, which we will analyse in the following sections.\\

Hawking proposed an explanation for the radiation emitted by black holes that has become very popular [6, 7]. Schematically, the explanation is as follows. The Heisenberg uncertainty principle allows empty space to fluctuate, creating pairs of virtual particles that suddenly appear, separate and come together to annihilate each other and disappear, being absorbed by the vacuum before they can be detected. In the presence of a black hole, one member of a pair of particles can cross its horizon inwards, leaving the other member free to escape to infinity. It will appear to a distant observer that the particles escaping from the black hole have been radiated by it. The black hole spectrum is exactly what one would expect from a hot body with a temperature $T_{H}$ that is inversely proportional to its mass, as shown in Eq. (2).\\

As with any hot body, a black hole emits mainly photons. However, if $T_{H}$ is sufficiently high or $M_{BH}$ is sufficiently small, it is possible that other particles such as neutrinos, electrons, protons, etc. may be emitted. Hawking radiation therefore implies a gradual reduction of $M_{BH}$ that can be thinked as an evaporation process.\\

It is possible to obtain an estimate of $T_{H}$ by means of a simple dimensional argument [8]. According to the Wien displacement law [9], the absolute temperature $T$ of a black body and the wavelength $\lambda_{max}$ at which the maximum emission occurs are related by the equation:

\begin{equation} 
\lambda_{max} T = \dfrac{2\pi\hbar c}{5k} = 2.90\times 10^{-3} m\cdot K.
\end{equation}

Since $R_{S}$ is the only characteristic length of the black hole, assuming that it behaves as a black body, we can take:

\begin{equation}	
\lambda _{max} \approx R_{S} = \dfrac{2GM_{BH}}{c^{2}}.
\end{equation}

Introducing this relation into Eq. (3) and solving for $T \approx T_{H}$, we obtain a result that differs from Eq. (2) only in the dimensionless constants:

\begin{equation}	
T_{H} \approx \dfrac{\hbar c^{3}}{(5/\pi)kGM_{BH}}.
\end{equation}

This equation reveals that $T_{H}$ is only significant for low-mass black holes. In his articles of 1974–5, Hawking speculated on the existence of this type of object, known as \textit{micro black holes} [4]. According to Hawking, the mass of a micro hole is less than $\sim 10^{12} kg$. If we enter this value into Eq. (1) we find that $R_{S} \sim 10^{-15} m$, which is the radius of a proton. If we take $M \sim 10^{12} kg$ in Eq. (2), we find that $T_{H} \sim 10^{11} K$, a very high temperature; according to Eq. (3), this has a maximum emission in the domain of gamma rays, $\lambda _{max} \sim 10^{-14} m$. Although this radiation is detectable, there is no astronomical evidence for micro holes emitting gamma rays. The least massive holes for which there is observational evidence are those of stellar mass, for which $M_{BH}$ is of the order of the solar mass, $\sim 10^{30} kg$. Taking $M_{BH} \sim 10^{30} kg$ in Eq. (2) results in $T_{H} \sim 10^{-7} K$, a temperature whose maximum emission is in the domain of long radio waves, $\lambda _{max} \sim 10^{4} m$, which is undetectable by astronomical observations [10].

\section{Bekenstein-Hawking entropy}

Since temperature implies the existence of entropy, then according to Hawking's calculations, a static black hole must have an entropy that depends on the area $A$ of the horizon:

\begin{equation}	
S_{BH} = \dfrac{kc^{3} A}{4\hbar G} = 1.33\times 10^{46} J/K \left(\dfrac{A}{m^{2}}\right), 
\end{equation}

where $A$ is calculated from Eq. (1):

\begin{equation}	
A = 4\pi R_{S} ^{2} = \dfrac{16\pi G^{2} M_{BH} ^{2}}{c^{4}}.
\end{equation}

The subscript $BH$ in Eq. (6) corresponds to the initials of Bekenstein-Hawking (although it could also be interpreted as “black hole”). The notion of black hole entropy was originally suggested in 1973 by Jacob Bekenstein [11], but it was Hawking who carried out a detailed calculation of $S_{BH}$ and understood its physical implications in greater depth.\\

We can perform a very simple dimensional calculation of $S_{BH}$, remembering that entropy is the energy per unit of temperature. For an observer at rest with respect to a static black hole, the total energy depends only on the mass:

\begin{equation}	
E_{BH} = M_{BH} c^{2}.
\end{equation}

It is therefore possible to estimate $S_{BH}$ as:

\begin{equation}	
S_{BH} \approx \dfrac{E_{BH}}{T_{H}} =\dfrac{M_{BH} c^{2}}{T_{H}} = \dfrac{8\pi kGM_{BH} ^{2}}{c\hbar}.
\end{equation}

Solving Eq. (7) for $M_{BH} ^{2}$ and introducing this value into Eq. (9), we obtain:

\begin{equation}	
S_{BH} \approx \frac{kc^{3} A}{2\hbar G}.
\end{equation}

This expression only differs by a factor of $1/2$ from Eq. (6) of Hawking. To verify that this equation does not violate the second law of thermodynamics, according to which the entropy of a closed system never decreases, we define the generalised entropy:

\begin{equation}	
S_{Gen} = S+S_{BH} = S+ \frac{kc^{3} A}{4\hbar G},
\end{equation}

where $S$ represents the total entropy of all the matter in the universe, outside of black holes. It is observed that although $S_{BH}$ and $A$ can increase or decrease individually, $S_{Gen}$ will never decrease. If a black hole absorbs material from its surroundings, $M_{BH}$ increases and therefore $A$ increases; however, at the same time, $S$ decreases, so that $S_{Gen}$ does not decrease. On the other hand, the emission of Hawking radiation decreases $M_{BH}$ and $A$, but increases $S$, again meaning that $S_{Gen}$ does not decrease.\\

Let us recall that in statistical mechanics, entropy is proportional to the logarithm of the number $W$ of microstates of a physical system that are compatible with a certain macrostate; in other words, the entropy is proportional to the number $W$ of configurations that the microscopic components of a system can adopt so that the system looks macroscopically identical. We can therefore estimate $S_{BH}$ as:

\begin{equation}	
S_{BH} = k \ln (W_{BH}) 
\end{equation}

If we eliminate $S_{BH}$ from Eqs. (10) and (12), we obtain:

\begin{equation}	
W_{BH} = e^{c^{3} A/4\hbar G} = e^{4\pi GM_{BH}	^{2} /\hbar c}.
\end{equation}

Since the macrostate of a static hole depends only on $M_{BH}$, we conclude that $W_{BH}$ is the number of microstates that are compatible with a given value of $M_{BH}$. Although no one knows exactly what these microstates represent, it is possible to prove that their number is colossal, meaning that the entropy of a black hole is huge. In effect, if we introduce into Eq. (13) the value of $M_{BH}$ for a stellar hole of $\sim 10^{30} kg$, we find that $W_{BH} \approx 10^{10^{77}}$. Black holes are the objects with the highest entropy in the universe.

\section{Evaporation time}

As mentioned in Section 2, Hawking radiation implies that a black hole evaporates gradually. In this section, we estimate the evaporation time, $t_{ev}$, which is defined as the time taken for the mass of a black hole to be reduced to zero. In his works of 1974 and 1975, Hawking made a crude estimate of $t_{ev}$ [4, 5]. More detailed calculations show that [12]:

\begin{equation}	
t_{ev} \approx 10^{-20} s \left(\frac{M_{BH}}{kg}\right)^{3}, 
\end{equation}

where $M_{BH}$ is the initial mass of the static black hole.\\ 

Next, we will develop a heuristic argument that allows for an approximate calculation of $t_{ev}$. For simplicity, we will assume that the Hawking radiation is composed only of photons.\\

According to Planck's law, the frequency $\nu$ and wavelength $\lambda$ of a photon are related to its energy $E$ by the expression:

\begin{equation}	
E = h\nu = 2\pi\hbar \nu = \frac{2\pi \hbar c}{\lambda}.
\end{equation}

From Eq. (4), we can take $\lambda \approx \lambda _{max} \approx R_{S}$: 

\begin{equation}	
E \approx \dfrac{\pi \hbar c^{3}}{GM_{BH}}.
\end{equation}

Since $R_{S}$ is the characteristic length of a black hole and $c$ is the speed of the emitted photons, dimensionally, the characteristic time for the emission of each photon will be:

\begin{equation}	
t_{c} \approx \frac{R_{S}}{c} =\frac{2GM_{BH}}{c^{3}}.
\end{equation}

Then, from Eqs. (15) and (17), the luminosity of a black hole can be estimated as:

\begin{equation}	
L_{BH} \approx \frac{E}{t_{c}} \approx \frac{\pi\hbar c^{6}}{2G^{2} M_{BH} ^{2}}.
\end{equation}

By definition, a black hole evaporates when its total energy $E_{BH}$ is emitted in the form of Hawking radiation. Consequently, the quotient between Eqs. (8) and (18) provides the evaporation time:

\begin{equation}	
t_{ev} \approx \frac{E_{BH}}{L_{BH}} =\frac{2G^{2} M_{BH} ^{3}}{\pi\hbar c^{4}} = 3.33\times 10^{-21} s \left(  \frac{M_{BH}}{kg}\right)^{3}.  
\end{equation}

This relationship differs from Eq. (14) only by a factor of 10. Eq. (14) reveals that for massive holes, $t_{ev}$ is extraordinarily large, meaning that evaporation is only detectable for small black holes. Indeed, if we introduce the mass of a micro-hole into Eq. (14), $M_{BH} \sim 10^{12} kg$, we find that $t_{ev} \sim 10^{16} s$, which is close to the age of the universe, $\sim 10^{17} s$; this implies that these micro holes should be completing their evaporation now, in the midst of large gamma-ray bursts. However, if we take a hole of stellar mass, $M_{BH} \sim 10^{30} kg$, this results in $t_{ev} \sim 10^{70} s$, a figure that is 53 orders of magnitude greater than the age of the universe. Since stellar black holes are the smallest for which there is observational evidence, as discussed in Section 2, we conclude that it is very difficult to detect the evaporation process.

\section{Final comments: Is there evidence for Hawking's theories?}

Since Hawking proposed his revolutionary ideas, astronomers have probed the space for evidence to back them up. Despite the efforts made, they have not observed black holes evaporating or emitting Hawking radiation [10]. On the other hand, the less massive black holes for which there is astronomical evidence are the stellar ones, whose quantum behavior, as we know, is completely undetectable\footnote{Researchers have also speculated on the possibility of generating micro black holes in particle accelerators such as the LHC (Large Hadron Collider), but there is no empirical evidence to support this idea.}. Does this mean that Hawking's findings are condemned to remain on a theoretical ground? The answer is no, since recently a new research field, unrelated to astronomy, has started testing some Hawking’s predictions in the laboratory.\\

The first physicist to propose that the empirical verification of Hawking's theories may not be in the stars, was William G. Unruh, who in 1981 suggested that the Hawking radiation could be observed by means of a \textit{black hole analogue model} [13]. An analogue model is a system designed to mimic the behavior of some object or physical system difficult to observe in its natural state. Therefore, the objective of the analogue model is to reproduce as closely as possible the physical behavior of the original phenomenon in order to understand its properties. In 2014, Jeff Steinhauer, from the physics department of the \textit{Israel Institute of Technology}, managed to create an analogous model such as the one suggested by Unruh [14]. The black-hole-analogue developed by Steinhauer allowed him to observe the emission of Hawking radiation, with a temperature given approximately by equation (2).\\

Recently, Steinhauer and his team have managed to extend and improve the results of the 2014 experiment [15]. Although the results obtained are not very accurate due to the noise inherent in the measurement process, the findings again agree with Hawking's prediction, at least within the precision with which the experiment allows to identify a temperature. Despite that the details of Steinhauer's experiment are complex, and his analysis goes beyond the scope of this work, what is important for our purposes is to point out that perhaps the corroboration of Hawking's theories is not found in the stars but on Earth. However, Steinhauer's experiment is inconclusive and needs to be confirmed by others.\\

We see then that there is still much work to be done to empirically corroborate Hawking's findings, so that in the short term it is difficult to obtain solid empirical results, either through black-hole-analogues or through astronomical observations. In any case, the importance of the contribution of the British genius is undoubted, and when the physical implications are fully understood, it will surely be considered one of the great scientific revolutions of the twentieth century.

\section*{Acknowledgments}
I would like to thank to Michael Van Sint Jan and Daniel Rojas-Líbano for their valuable comments in the writing of this paper. 

\section*{References}

[1] K. Ferguson, Stephen Hawking. Su vida y obra, Crítica, Buenos Aires, 2012.

\vspace{2mm}
[2] A. Einstein, Die Grundlage der allgemeinen Relativitätstheorie, Annalen der Physik, 354 (1916) 769-822.

\vspace{2mm}
[3] J. Pinochet, The Hawking temperature, the uncertainty principle and quantum black holes, Phys. Educ., 53 (2018) 1-6.

\vspace{2mm}
[4] S.W. Hawking, Black Hole explosions?, Nature, 248 (1974) 30-31.

\vspace{2mm}
[5] S.W. Hawking, Particle creation by black holes, Communications in Mathematical Physics, 43 (1975) 199-220.

\vspace{2mm}
[6] S.W. Hawking, The Universe in a Nutshell, Bantam Books, London, 2001.

\vspace{2mm}
[7] S.W. Hawking, A brief history of time, Bantam Books, New York, 1998.

\vspace{2mm}
[8] J. Pinochet, Hawking Temperature: An elementary approach based on Newtonian Mechanics and Quantum Theory, Physics Education, 51 (2016) 1-6.

\vspace{2mm}
[9] P.A. Tipler, R.A. Llewellyn, Modern Physics, 6 ed., W. H. Freeman and Company, New York, 2012.

\vspace{2mm}
[10] M.L. Kutner, Astronomy:A Physical Perspective, 2 ed., Cambridge University Press, New York, 2003.

\vspace{2mm}
[11] J.D. Bekenstein, Black Holes and Entropy, Physical Review D, 7 (1973) 2333-2346.

\vspace{2mm}
[12] V.P. Frolov, A. Zelnikov, Introduction to Black Hole Physics, Oxford University Press, Oxford, 2011.

\vspace{2mm}
[13] W.G. Unruh, Experimental Black-Hole Evaporation?, Physical Review Letters, 46 (1981) 1351-1353.

\vspace{2mm}
[14] J. Steinhauer, Observation of self-amplifying Hawking radiation in an analogue black-hole laser, Nature Physics, 10 (2014) 864-869.

\vspace{2mm}
[15] J.R. Munoz de Nova, K. Golubkov, V.I. Kolobov, J. Steinhauer, Observation of thermal Hawking radiation at the Hawking temperature in an analogue black hole, arXiv:1809.00913v2 [gr-qc] (2018).

\end{document}